\documentclass{article}
\usepackage{spconf,amsmath,graphicx}
\usepackage{algorithmic}
\usepackage{algorithm}
\usepackage{multirow}
\usepackage{amsmath,amssymb}
\usepackage{graphicx}
\usepackage[dvipsnames]{xcolor}

\title{Cross-modal Knowledge Distillation for Vision-to-Sensor Action Recognition}
\name{Jianyuan Ni$^{1}$\qquad Raunak Sarbajna$^{2}$ \qquad Yang Liu $^{3}$\qquad Anne H.H. Ngu$^{1}$ \qquad Yan Yan $^{4}$}

\address{$^{1}$ Texas State University, USA \\
      $^{2}$ University of Houston, USA \\
      $^{3}$ Sun-Yat-Sen University, China \\
      $^{4}$ Illinois Institute of Technology, USA}

\begin{document}
%
\maketitle
\begin{abstract}
Human activity recognition (HAR) based on multi-modal approach has been recently shown to improve the accuracy performance of HAR. However, restricted computational resources associated with wearable devices, \emph{i.e.,} smartwatch, failed to directly support such advanced methods. To tackle this issue, this study introduces an end-to-end Vision-to-Sensor Knowledge Distillation (\emph{VSKD}) framework. In this \emph{VSKD} framework, only time-series data, \emph{i.e.,} accelerometer data, is needed from wearable devices during the testing phase. Therefore, this framework will not only reduce the computational demands on edge devices, but also produce a learning model that closely matches the performance of the computational expensive multi-modal approach. In order to retain the local temporal relationship and facilitate visual deep learning models, we first convert time-series data to two-dimensional images by applying the Gramian Angular Field (\emph{GAF}) based encoding method. We adopted ResNet18 and multi-scale TRN with BN-Inception as teacher and student network in this study, respectively. A novel loss function, named Distance and Angle-wised Semantic Knowledge loss (DASK), is proposed to mitigate the modality variations between the vision and the sensor domain. Extensive experimental results on UTD-MHAD, MMAct and Berkeley-MHAD datasets demonstrate the effectiveness and competitiveness of the proposed VSKD model which can deployed on wearable sensors.
\end{abstract}
\section{Introduction}
\label{sec:intro}
Human Activity Recognition (HAR), \emph{i.e.,} perceiving and recognizing human actions, is crucial for real-time applications, such as healthcare and human-robot interaction \cite{vrigkas2015review}. Vision-based methods have dominated the HAR community because video data contains rich appearance information \cite{sun2020human}. However, video-based HAR is intrinsically restricted in various occlusion and illumination conditions similar to the human vision limitations. Meanwhile, utilizing time-series data, \emph{i.e.,} accelerometer data, from wearable devices is another typical way of identifying the HAR problem. But sensor-based HAR approaches are difficult to achieve reliable performance compared to video modality due to the constraint of single context information  \cite{guo2016wearable, mauldin2018smartfall}. For instance, previous work indicated that deep learning method using accelerometer data from a wrist-worn watch for fall detection only achieved 86\% accuracy due to the fact that it is difficult to differentiate various wrist movements when someone falls \cite{mauldin2018smartfall}. Multi-modal HAR systems can solve such problems by utilizing the complementary information acquired from different modalities. For example, vision-based approaches could provide global motion features while sensor-based methods can give  3D  information  about  local  body  movement \cite{liu2021semantics}. Nevertheless, limited resources associated with the wearable devices, such as CPU and memory storage, cannot support such powerful and advanced multi-modal systems. In order to overcome such issues, the technique of cross-modal transfer, \emph{i.e.,} knowledge distillation (KD), that
needs only one modality input during the testing phase to reach the performance close to the combination of multi-modal data during the training phase has been proposed \cite{hinton2015distilling}.
Using this approach, we can transfer the knowledge from vision to sensor domain by reducing computation resource demand, but also eventually boost the performance of HAR using wearable devices.

\begin{figure} [t]
 \center
  \includegraphics[width=8.1 cm]{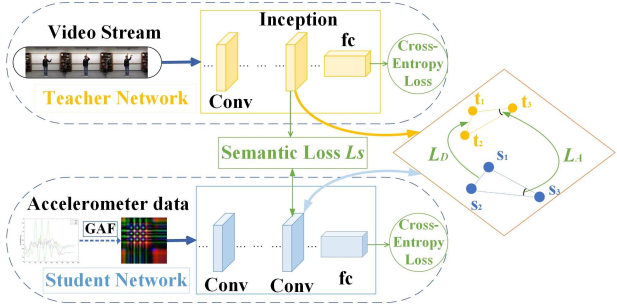}
  \caption{Schematic overview of the proposed \emph{VSKD} method.} 
  \vspace{-0.5 cm}
  \label{figure:framework}
\end{figure}

In this paper, we propose an end-to-end Vision-to-Sensor Knowledge Distillation (VSKD) for HAR recognition. The overview of the proposed method is shown in Figure \ref{figure:framework}. First, we adopted the Gramian Angular Field (\emph{GAF}) to encode the accelerometer data to an image representation while preserving 
the temporal information from the accelerometer data \cite{setiawan2019deep}. Next, we trained the teacher networks with video stream inputs with the standard cross-entropy loss function. The KD process of accelerometer data was accomplished by using the new loss function, named Distance and Angle-wised Semantic Knowledge loss (\emph{DASK}). Overall, the contributions of this paper are summarized as follows: 1) To the best of our knowledge, this is the first study conducting the knowledge distillation (KD) model from the video-to-sensor domain. In this \emph {VSKD} model, a student network with input of wearable sensor data, \emph{i.e.,} accelerometer data, which learns the compensatory information from the teacher network with input of video streams. 2) We proposed a novel loss function (\emph{DASK}), which is utilized to alleviate the modality gap between the teacher and student network. 3) We demonstrated the effectiveness and robustness of the proposed \emph {VSKD} method on three public datasets.
\section{Related Work}
HAR has been an active research field due to its wide application in various areas
\cite{vrigkas2015review,sun2020human, wang2021m}. Despite the fact that video modality containing rich RGB information, video modality is subject to various viewpoints or illumination conditions which affects its effectiveness. HAR studies with time-series data, \emph{i.e.,} accelerometer data, from wearable devices is growing rapidly \cite{mauldin2018smartfall,ahmad2019multidomain, karagiannaki2017online}. Although those works demonstrated the feasibility of sensor-based HAR approaches, they cannot achieve reliable performance due to the  noisy data or sensor variations \cite{guo2016wearable}. By aggregating the advantages of various data modalities, a multi-modal approach can ultimately provide a robust and accurate HAR method. However, the limited computation capabilities of a low-cost wearable devices prevent the complexity of multi-modal methods that can be deployed on the device directly. In order to build lightweight and accurate HAR models, the knowledge distillation approach has emerged to build a student model with less computational overhead and yet can retain similar accuracy performance as the teacher model \cite{hinton2015distilling}. For example, Kong {\emph{et al.}} \cite{kong2019mmact} proposed a multi-modal attention distillation method to model video-based HAR with the instructive side information from inertial sensor modality. Similarly, Liu {\emph{et al.}} \cite{liu2021semantics} introduced a multi-modal KD method where the knowledge from multiple sensor data were adaptively transferred to video domain. Even though those works provide promising results on HAR with the multi-modal approach, no multi-modal KD work has yet been proposed where the time-series data is used as the student model. Using the reversed approach will improve the accuracy performance of a sensor-based HAR, but also reduce the computational resource demand making it viable to run the model on the wearable devices directly. 

\section{Methodology}
\begin{figure}
 \center
  \includegraphics[width=8.8 cm]{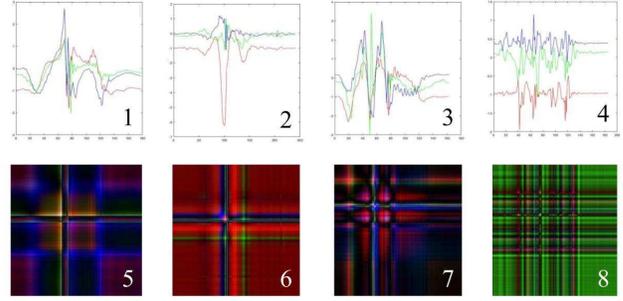}
  \caption{Selected sensor (top) and their corresponding GAF images (bottom) in UTD-MHAD \cite{chen2015utd} : (1) basketball shooting; (2) bowling; (3) knock on door and (4) walking.}
  \vspace{-0.5 cm}
  \label{figure:GAF image}
\end{figure}
\subsection{Virtual Image Generation}

Inspired by \cite{setiawan2019deep}, we encodes the accelerometer data to image representation first. In short, we denote one of the three axial accelerometer data (for example, $x$ coordinate) as $ \mathbf {X} = \lbrace x_1, x_2, ..., x_n\rbrace $ and normalize it into $ \mathbf{\hat{X}} $ among interval [-1, 1]. The normalized $ \mathbf{\hat{X}} $ was then encoded into the polar coordinate $ \left( \theta,\gamma \right) $ from the normalised amplitude and the radius from the time $t$, as represented in Eq.\ref{eq:1}:
\begin{equation}
\small
\centering
\begin{aligned}
\mathbf {\emph {g}} \left( \hat{x_i}, t_i \right) = [\theta_i, r_i]  
  \quad \textrm{where} \quad
        \begin{cases}
            \theta_i = arccos (\hat{x_i}),  x_i \in  \mathbf{\hat{X}} \\
            r_i = t_i \\ 
        \end{cases}
\end{aligned}
\label{eq:1}
\end{equation}
After this transformation, the correlation coefficient between vectors can be easily calculated using the trigonometric sum between points \cite{setiawan2019deep}. The tri-axial sensor data with the size of $n$ can be assembled as an image representation $\mathbf {P} = \left( \mathbf {G_x}, \mathbf {G_y}, \mathbf {G_z} \right) $ of size $n\times n \times3$. Selected examples of sensor and their corresponding GAF images in UTD-MHAD \cite{chen2015utd} are shown in Figure \ref{figure:GAF image}. 

\subsection{ DASK Loss}

Hinton {\emph{et al.}} \cite{hinton2015distilling} proposed a KD method which compress knowledge from larger mode (i,e.  \emph{teacher}) into a smaller model (i,e. \emph {student}), while retaining decent accuracy performance. Given a teacher model \emph {$T_k$} and a student model \emph{$S_k$}, the soft-target $\tilde {y}^T$ produced by the teacher model is considered as high-level knowledge. The loss of KD when training student model can be defined as:
\begin{equation}
\small
\begin{aligned}
\mathcal{L}_{KD} = \mathcal{L_C}(y,y^S) + \alpha \mathcal L_K (\tilde{y}^T, \tilde{y}^S)
\end{aligned}
\label{eq:2}
\end{equation}
\begin{equation}
\small
\begin{aligned}
 \mathcal L_K = \frac{1}{m} \sum_{k=0}^ m KL ({\frac {{P}^{T_k}} {T} }, {\frac {P^{S_k}} {T} }) 
\end{aligned}
\label{eq:3}
\end{equation}
where \emph{y} and $y^S$ refer to the predicted labels and class probability for the student network in this study, respectively. $\tilde{y}^S$ is the soft target generated by the student model. Here $\mathcal L_C $ is the typical cross-entropy loss and $\mathcal L_K$ is the Kullback-Leibler (KL) divergence, while $ P^{T_k} $ is the class probability for the teacher network and $ P^{S_k} $ is the class probability for the student network. \emph{T} represents the temperature controlling the distribution of the provability and we use \emph{T} = 4 in this study according to \cite{hinton2015distilling}.

KD methods \cite{hinton2015distilling, park2019relational} assume the knowledge as a learned mapping from inputs to outputs, which means the outputs themselves contain some relative information from inputs. Therefore, in order to minimize the modality gap between the vision and the sensor domain, we focus on information transfer. More specifically, not only do we to conduct the VSKD based on individual predicted outputs,  we also need to consider the structural relation, such as the distance and the angle information, as well as the semantic information among those two modalities that share the same action activity. Therefore, given a pair of training examples, the distance-wise function $\mathcal \psi_D$ tries to minimize the Euclidean distance between teacher and student examples. $\mu$ is a normalization factor for distance and  $l_\delta$ is Huber loss. The distance-wise distillation loss ${L_D}$, which tries to penalize the distance differences between teacher and student outputs is defined as:
\begin{equation}
\small
\begin{aligned}
\mathcal{\psi}_D (t_i, t_j,t_k) =  \frac{1}{\mu} \lVert \mathcal (t_i - t_j) \rVert_2
\end{aligned}
\label{eq:4}
\end{equation}
\begin{equation}
\small
\begin{aligned}
 \mathcal L_D = \sum_{(x_i,x_j) \in X^2} l_\delta ({\mathcal{\psi}_D (t_i, t_j) }, {\mathcal{\psi}_D (s_i, s_j) }) 
\end{aligned}
\label{eq:5}
\end{equation}
Similarly, given three training examples, the angle-wise function $\mathcal \psi_A $ tries to minimize the angle between teacher and student examples. The angle-wise distillation loss ${L_A}$  which tries to transfer the angle-relation information among teacher and students outputs is defined as:
\begin{equation}
\small
\begin{aligned}
\mathcal{\psi}_A (t_i, t_j) = \cos \angle {}t_j t_j t_k = \langle {e^i}j,{e^k}j \rangle
\\ \textrm{where} \hspace{1cm}
{e^i}^j = \frac{t_i-t_j} {\lVert \mathcal (t_i - t_j) \rVert_2}, {e^k}^j = \frac{t_k-t_j} {\lVert \mathcal (t_k - t_j) \rVert_2}
\end{aligned}
\label{eq:6}
\end{equation}
\begin{equation}
\small
\begin{aligned}
 \mathcal L_A = \sum_{(x_i,x_j,x_k) \in X^2} l_\delta ({\mathcal{\psi}_A (t_i, t_j,t_k) }, {\mathcal{\psi}_A (s_i, s_j,s_k) }) 
\end{aligned}
\label{eq:7}
\end{equation}
In addition, since multi-modal data have the same semantic content, semantic loss is defined as:
\begin{equation}
\small
\begin{aligned}
 \mathcal L_S = \frac{1}{m} \sum_{k=1}^ m (\|H^S - H^T\|)_2^2 
\end{aligned}
\label{eq:8}
\end{equation}
where $H^S$ and $H^T$ represents the feature of the layer prior to the last fc layer, respectively.

In summary, we use the original KD loss $ L_{K_D}$ and augment it to include the distance and the angle-wised distillation loss $ L_D, L_A$ as well as the semantic loss $L_S$, to train the student network and the final DASK loss for the student model is defined as follow:
\begin{equation}
\small
\begin{aligned}
 \mathcal L_T^S = L_{KD} + \beta (L_D +  L_A) + \gamma L_S
\end{aligned}
\label{eq:9}
\end{equation}
where $\alpha, \beta, \gamma $ are the tunable hyperparameters to balance the loss terms for the student network.  

\section{Experiments}

\subsection{Dataset}

In this study, three benchmark datasets were selected due to their multi-modal data forms: \emph{MMAct} \cite{kong2019mmact}, \emph{UTD-MHAD} \cite{chen2015utd} and \emph{Berkeley-MHAD} \cite{ofli2013berkeley}. We use video streams as the teacher modality and accelerometer data as the student modality in those datasets.
\begin{table}[t]
\begin{center}
\scalebox{0.90}{
\begin{tabular}{|c|c|c|}
\hline
 Method & Testing Modality & Accuracy (\%)\\
 \hline
Singh  {\emph{et al.}} \cite{singh2020deep} & Acc. + Gyro. & 91.40 \\
\hline
Ahmad and Khan \cite{ahmad2019multidomain}  & Acc. + Gyro. &  95.80\\
\hline
Wei  {\emph{et al.}} \cite{wei2019fusion}& Acc. + Gyro. & 90.30  \\
\hline
Chen  {\emph{et al.}} \cite{chen2016fusion} & Acc. + Gyro. & 96.70  \\
\hline
Garcia-Ceja  {\emph{et al.}} \cite{garcia2018multi} & Acc. & 90.20  \\
\hline
Student baseline & Acc. & 94.87 \\
\hline
\bf VSKD model & \bf Acc. & \bf 96.97 (2.1 $\uparrow$) \\
\hline
\end{tabular}}
\end{center}
\caption{Comparison result on accuracy performance of UTD-MHAD. The number in parenthesis means increased accuracy  over the student baseline. Acc. denotes accelerometer and Gyro. denotes gyroscope.}
\vspace{-0.1 cm}
\label{table:1}
\end{table} 
\begin{table}[t]
\begin{center}
\scalebox{0.75}{
\begin{tabular}{|c|c|c|c|}
\hline
 Method & Testing Modality & Accuracy (\%) & F1 score (\%)\\
\hline
Das  {\emph{et al.}} \cite{das2020mmhar} & Acc.(Six locations) & 88.90 & 88.80\\
\hline
Student Baseline & Acc. (Left Wrist) & 89.09 & 89.27\\
\hline
\bf VSKD model & \bf Acc. (Left Wrist) & \bf 90.18 ( 1.09 $\uparrow$ ) & \bf 91.58\\
\hline
Student Baseline & Acc. (Right Wrist) & 86.54 & 88.33\\
\hline
VSKD model & Acc. (Right Wrist) &   87.64 (1.10 $\uparrow$ ) & 89.90\\
\hline
Student Baseline & Acc. (Left Hip) & 83.27 & 84.45\\
\hline
VSKD model & Acc. (Left Hip) & 83.82 ( 0.55 $\uparrow$ ) & 83.91\\
\hline
Student Baseline & Acc. (Right Hip) & 81.09  & 81.45\\
\hline
VSKD model & Acc. (Right Hip) &  82.55 (1.46 $\uparrow$ ) & 82.99\\
\hline
Student Baseline & Acc. (Left Ankle) & 65.09 & 64.73\\
\hline
VSKD model &  Acc. (Left Ankle) & 65.82 ( 0.73 $\uparrow$ ) & 66.90\\
\hline
Student Baseline & Acc. (Right Ankle) & 62.91 & 62.73\\
\hline
VSKD model & Acc. (Right Ankle) & 64.36 ( 1.45 $\uparrow$ ) & 63.60\\
\hline
\end{tabular}}
\end{center}
\caption{Comparison result on accuracy and F1 performance of Berkeley-MHAD. The number in parenthesis means increased accuracy and F1 score over the student baseline.}
\vspace{-0.4 cm}
\label{table:2}
\end{table}  
\subsection{Experimental Settings}

For the teacher network, we used multi-scale TRN \cite{zhou2018temporal} with BN-Inception pre-trained on ImageNet due to its balance between the number of parameters and efficiency. In the teacher network, we set the dropout ratio as 0.5 to reduce the effect of over-fitting. The number of segments is set as 8 for Berkeley-MHAD and UTD-MHAD, while 3 for the MMAct. For the student baseline model, we used ResNet18 as the backbone. All the experiments were performed  on four Nvidia GeForce GTX 1080 Ti GPUs using PyTorch. We employed the classification accuracy and F-measure as the evaluation metric to compare the performance of the VSKD model with: 1) a student baseline model (ResNet18); 2) other work in which time-series data were applied.

\begin{table}[t]
\begin{center}
\scalebox{0.70}{
\begin{tabular}{|c|c|c|c|}
\hline
 Method & Testing Modality & Cross Subject (\%) & Cross Session (\%) \\
\hline
Kong  {\emph{et al.}} \cite{kong2019mmact} & Acc.(Watch+Phone) &  62.67 &   70.53  \\
\hline
Kong  {\emph{et al.}} \cite{kong2020cycle} & RGB video &  65.10 &   62.80  \\
\hline
Student baseline & Acc. (Phone) & 55.44 &  61.38 \\
\hline
\bf VSKD model & \bf Acc. (Phone) & \bf 65.83  (  10.39 $\uparrow$ ) & \bf 73.64 (  12.26 $\uparrow$ )\\
\hline
Student baseline & Acc. (Watch) & 46.83 & 20.63 \\
\hline
VSKD model & Acc. (Watch) & 60.14 (  13.31 $\uparrow$ ) & 40.82 (  20.19 $\uparrow$ ) \\
\hline
\end{tabular}}
\end{center}
\caption{Comparison result on F-measurement performance of MMAct. The number in parenthesis means increased F1 score over the student baseline.}
\vspace{-0.5 cm}
\label{table:3}
\end{table}
\subsection{Experimental Results}

The comparison results of three datasets are shown in Table \ref{table:1}, \ref{table:2}, and \ref{table:3}, respectively. In Table \ref{table:1}, the proposed VSKD model performs better than all the previous comparable models. We make an improvement in the testing accuracy of 6.77\% compared to the accelerometer view method which extracted 16 features from accelerometer signals for classification \cite{garcia2018multi}. The VSKD model achieved 2.1\% higher in accuracy performance, compared to just the student model alone. This result sheds light on incorporating video modality for improving sensor-based HAR. It is worth noting that the proposed VSKD model even performs better as compared to the methods where the accelerometer and gyroscope data were used for testing \cite{singh2020deep, ahmad2019multidomain, wei2019fusion, chen2016fusion}. These results demonstrated that accelerometer data in the VSKD model can significantly learn knowledge from video streams and thus making an improvement in testing accuracy by 0.37\%-6.77\%. In Table \ref{table:2}, the proposed VSKD model trained with vision and sensor modality can outperform all the student baseline models. Even though gray-scale video streams on the Berkeley-MHAD dataset lack color information which may degrade the knowledge transfer process, the improvements ranged from 0.55\% to 1.46\% can still be  obtained by the additional support of multi-modal modalities. Also, the VSKD model tested with the left wrist acceleromeer data performs better compared to the previous study where accelerometer data from six locations were used \cite{das2020mmhar}. In Table \ref{table:3}, while accelerometer data from the phone is the only modality in the testing phase, the method achieves better F-score performance compared to \cite{kong2019mmact,kong2020cycle} in which either video streams or accelerometer data from phone and watch was used in the testing phase. This validates that the VSKD approach can effectively learn knowledge from the video modality to improve the accuracy performance of sensor-based HAR. We also note that the VSKD method trained with accelerometer data from the watch performs worse than the one with accelerometer data from phone. This result was consistent with previous works which showed arm movements introduce additional variability giving rise to a degradation in HAR \cite{san2018robust}.

\subsection{Ablation Study}

To evaluate the contribution of the proposed DASK loss function, we compare the DASK function with previous KD methods \cite{hinton2015distilling, zagoruyko2016paying,tung2019similarity}. For those KD methods, we use the shared codes, and the parameters are selected according to the default setting. As shown in Table \ref{table:4}, the proposed DASK loss function performs better than all of the previous comparable KD loss functions, proving that both structural relation and semantic information are critical information for time-series data in a KD process. Also, angle-wised loss ${L_A}$ contributes more (0.22\%) to accuracy improvement as compared to distance-wised loss ${L_D}$, indicating time-series data are more valuable in giving 3D information about local body movement. Furthermore, compared to structural relation information, semantic loss ${L_S}$ contributes more to accuracy improvement (0.70\%), which highlights the critical role of semantic information on sensor-based HAR. The proposed VSKD model with ResNet18 as the student baseline performed better (an accuracy of 96.97\% ) as compared to the VSKD model where VGG16 was used as the student baseline (95.34\%). This happens because different convolutional layers in ResNet tend to learn different types of features regarding the input \cite{prakash2021multi}, therefore, ResNet18 model is more effective in capturing these features from student model compared to VGG16 model. 
\begin{table}[t]
\begin{center}
\scalebox{0.75}{
\begin{tabular}{|c|c|c|c|}
\hline
Method  & Modality &  Accuracy (\%) & F1 score (\%) \\
\hline
ST \cite{hinton2015distilling} & Acc. &  96.04  & 96.15 \\
\hline
AT \cite{zagoruyko2016paying} & Acc. &  96.03 & 95.80\\
\hline
SP \cite{tung2019similarity} & Acc. &  95.80  & 95.57\\
\hline
DASK-VGG16 &  Acc. &   95.34  & 95.69 \\
\hline
DASK-ResNet18 &  Acc. &   \bf 96.97  & \bf 96.38\\
\hline
ASK (W/O D)-ResNet18 &  Acc. &   96.73  & 96.27\\
\hline
DSK (W/O A)-ResNet18 &  Acc. &    96.51  & 95.80\\
\hline
SK (W/O D and A)-ResNet18 &  Acc. &    96.50  & 96.06\\
\hline
DAK (W/O S)-ResNet18 &  Acc. &    95.80  & 96.04 \\
\hline
\end{tabular}}
\end{center}
\caption{Ablation study of accuracy and F1 score performance on UTD-MHAD dataset. W/O denotes Without. D denotes the distance-wise loss ${L_D}$. A denotes the angle-wise loss ${L_A}$. S denotes the semantic distillation loss ${L_S}$. }
\vspace{-0.4 cm}
\label{table:4}
\end{table}
\section{Conclusion}

In this paper, we propose an end-to-end Vision-to-Sensor Knowledge Distillation (\emph{VSKD}) model, which not only improve the sensor-based HAR performance, but also reduce the computational resource demand during the testing phase. We also propose a novel loss function (\emph{DASK}), which highlights the importance of structural relation and semantic information for bridging the modality gap between vision and sensor domain. Extensive experimental results on UTD-MHAD, MMAct and Berkeley-MHAD datasets demonstrate the effectiveness and competitiveness of the proposed VSKD model.

\vfill\pagebreak
\bibliographystyle{IEEEbib}
\bibliography{strings,refs}

\end{document}